\documentclass[prb,amsmath,twocolumn,showkeys,showpacs]{revtex4}
\usepackage{graphicx}
\usepackage{dcolumn}
\usepackage{amssymb}
\usepackage{bm}% bold math
\begin{document}
%%%%%%%%%%%%%%%%%%%%%%%%%%%%
\newcommand{\be}{\begin{equation}}
\newcommand{\ee}{\end{equation}}
\newcommand{\vk}{\mathbf{k}}
\newcommand{\vq}{\mathbf{q}}
\newcommand{\vp}{\mathbf{p}}
\newcommand{\vecr}{\mathbf{r}}
\newcommand{\vx}{\mathbf{x}}
\newcommand{\caH}{\mathcal{H}}
\newcommand{\la}{\langle}
\newcommand{\ra}{\rangle}
\newcommand{\e}{\epsilon}
\newcommand{\half}{\frac{1}{2}}
%%%%%%%%%%%%%%%%%%%%%%%%%%%%%%%%%%%
\title{ X-ray edge problem of  graphene}
\author{ S.-R. Eric Yang}
\affiliation{Department of Physics,
 Korea University, Seoul, Korea}
\author{Hyun C. Lee}
\email[the author to whom the correspondences should be addressed:~]{ hyunlee@sogang.ac.kr}
\affiliation{Department of Physics and Basic Science Research Institute,
 Sogang University, Seoul, Korea}
\date{\today}
%%%%%%%%%%%%%%%%%%%%%%%%%%%%%%%%%%%%%%%%%%%%%%%%%%%%%%%%
\begin{abstract}
The X-ray edge problem of  graphene with  the Dirac fermion spectrum  is studied.
At half-filling the linear density of states suppresses the singular response of the Fermi liquid, 
while away from half-filling the singular features of the Fermi liquid reappear. 
The crossover behavior as a function of the Fermi energy  is examined in detail.
The exponent of the power-law absorption rate depends both on the intra- and inter-valley scattering, and it 
changes  as a function of the Fermi energy, which may be tested experimentally.
\end{abstract}
%%%%%%%%%%%%%%%%%%%%%%%%%%%%%%%%%%%%%%%%%%%%%%%%%%%%%%%%%%%
\maketitle
%%%%%%%%%%%%%%%%%%%%%%%%%%%%%%%%%%%%%%%%%%%%%%%%%%%%%%%%%%%%%%%%%%
\section{Introduction} Recently the graphenes are being studied intensively mainly due to its novel
transport properties such as quantum Hall effect.\cite{exp1,exp2,exp3,exp4}
Many of such novel properties are attributed to the  Dirac fermion band structure of a graphene.\cite{Slo,ando1}
 The density of states (Dos) is linear in energy, 
and  vanishes at the Fermi energy at half-filling. Therefore, many properties of graphene
at half-filling are expected to be markedly different from those of the ordinary Fermi liquid (FL) which
has a finite Dos at the Fermi energy.
However, away from half-filling, the Dos of graphene does not vanish at the Fermi energy 
although it can be very small, and some   features of FL  are expected to emerge.
\begin{figure}[!hbt]
\begin{center}
\includegraphics[width = 0.45 \textwidth]{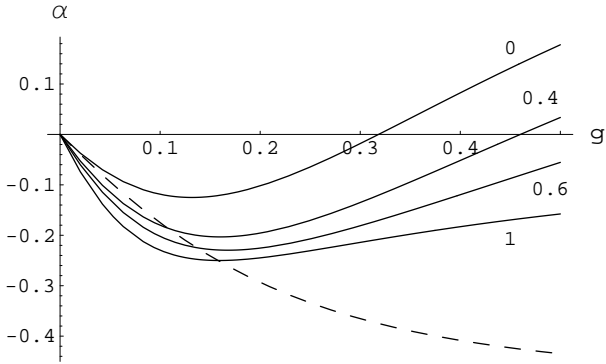}
\caption{The dependence of power-law exponent $\alpha$ of graphene on the coupling constant $g=\rho V_0$ for $N_c =8$
(see text for definitions). 
From top to bottom, $ \frac{V_1}{V_0} = 0, 0.4, 0.6, 1$.  Note that the dependence on $g$ implies the 
dependence on the the Fermi energy.
The dashed  curve is for the Fermi liquid.}
\label{expo}
\end{center}
\end{figure}

The problem of the transient optical response of FL has a long and interesting history.\cite{mahan1,mahan2,anderson,nozieres,schotte} 
The high-energy X-ray creates a deep core hole, 
and as a result the Fermi sea of conduction electrons reacts to a \textit{time-dependent} localized potential.
In terms of the optical response the conduction electrons react in two distinctive ways:
the excitonic processes \cite{mahan2} and the orthogonality catastrophe (OC) \cite{anderson}.
For an attractive potential  stemming from the created hole, the excitonic processes tend to enhance the optical 
response, while OC effects suppress the optical response for either repulsive or attractive potential.
It is crucial to include these two effects on equal footing.  For Fermi liquids with a short-range core hole potential $V(\vec{r})=-V\delta(r)$,
the absorption intensity which is asymptotically exact near threshold  takes the following form. \cite{nozieres} 
\be
\label{Noz}
I(\omega)\sim  \theta(\omega -\omega_T^*)\,\rho \, \left(\frac{\omega-\omega_T^*}{E_c} \right )^{\alpha_{\mathrm{FL}}},
\ee
where $\omega_T^*$ is the threshold energy which is the energy gap between the core level and the Fermi energy of FL with  corrections from particle-hole interaction. $\theta(\omega)$ is the step function.
The power-law exponent is given by\cite{comment}
\be
\label{exp}
\alpha_{{\rm FL}}=-2\frac{\delta}{\pi}+ N_s\Big(\frac{\delta}{\pi}\Big)^2,\quad
\delta=\tan^{-1}(\pi \rho V).
\ee
The constants $\rho$,  $E_c$,  $N_s$, $\delta$ are 
the Dos at the Fermi energy, a high-energy cutoff, the spin degeneracy $2$, and the scattering phase shift, respectively.

The X-ray edge problem and the  OC have the universal character
of the dynamic  quantum impurity  in gapless host system, 
and have been studied in diverse physical systems.
The tunneling junction with a localized impurity having multiple scattering channel  has been studied by
Matveev and Larkin \cite{larkin} using an analogy with the X-ray edge problem.
The X-ray edge problem and the OC in one-dimensional Luttinger liquid 
have been studied extensively in Ref.[\onlinecite{1dLL1,1dLL2,1dLL3,1dLL4,1dLL5}].
Khveshchenko and Anderson \cite{dima} studied the X-ray edge problem of the tomographic Luttinger liquid
which has a vanishing Dos at Fermi energy.
Patton and Geller \cite{PattonGeller} have obtained the 
anomalous tunneling Dos of  strongly correlated electron systems by mapping
 the original Hamiltonian to the one to  which  the method of X-ray edge problem can be applied.
We also note that the Kondo physics belongs to the same class  as the X-ray edge problem. 
A subclass of the Kondo physics problems where
the conduction electrons have power law Dos (soft gap) at Fermi energy  has been studied very extensively
mainly in the context of the anisotropic superconductors and the dynamical mean field theory.\cite{kondo1,kondo2,kondo3}

At a sufficiently large filling  graphene is expected to exhibit
FL characters.
On the other hand, at and near half-filling the conventional approaches to the optical response of FL cannot be applied directly 
due to the qualitative differences in band structure between graphene and FL.
For example, the bosonization method \cite{schotte,bosobook}
cannot be applied because the local Green function  of graphene cannot be represented by that of one-dimensional chiral fermions.\cite{bosobook}
Another new feature of graphene, not present in the ordinary Fermi liquid, is  the inter-valley scattering.  
Here we employ the method of functional integrals to overcome these problems, and 
integrate out all conduction electron degrees of freedom except for the one at the 
position of the immobile hole. 
As far as the local response is concerned, the resulting \textit{local action} contains all the necessary
information.\cite{kane,nagaosa}
The analysis of the local action based on path integral approach
enables us to study the FL and the graphene in a parallel way, highlighting the similarities and the differences
at the same time.  

In this paper we report theoretical results on the X-ray edge problem of graphene 
based on the massless Dirac fermion model.
We find that the X-ray absorption intensity has the following power-law dependence ($\omega > \omega_T^*$)
\be
\label{Noz2}
I(\omega)\sim \rho\,\left[  \Big(\frac{\omega-\omega_T^*}{E_c} \Big)^{\alpha}+ \frac{E_c}{\mu}\Big(\frac{\omega-\omega_T^*}{E_c}\Big)^{1+N_c (\delta/\pi)^2} \right ]
\ee
with the exponent 
\be
\label{exp2}
\alpha=-2\frac{\delta_+}{\pi}+N_c(\delta/\pi)^2, \quad
\delta^2 = \frac{1}{2} ( \delta_+^2 + \delta_-^2),
\ee
where 
$\delta_\pm = \tan^{-1} [\pi \rho ( V_0 \pm V_1) ]$ with the short-range intra- and inter-valley  scattering strengths
$ V_0$ and $ V_1$, respectively. In the basis of Dirac fermions (see below), the scattering potential takes the form of 
the following 4x4 matrix which is consistent with the symmetry of 
graphene.\cite{ando2}
\be
\label{potential}
V_{\mathrm{scat}}(\vec{r}) = \delta(\vec{r})\, \Big ( V_0 \,{\rm I}_4 +V_1 \sigma_x \otimes \tau_x \Big )
\equiv \delta (\vec{r}) \tilde{V}. 
\ee
The combination $V_0 \pm V_1$  in the phase shift $\delta_\pm$ is nothing but the eigenvalue of 
the scattering matrix $\tilde{V}$, and  the Green functions are also naturally decomposed with the corresponding eigenvectors,
see Eq.(\ref{ND}).
The exponent $\alpha$ of the first term  in Eq.(\ref{Noz2}) can be negative, which implies that  the absorption intensity
is singular at the threshold energy $\omega_T^*$. On the other hand, the second term is non-singular since  the corresponding
exponent  is always positive.
However, near half filling the second term of Eq.(\ref{Noz2}) dominates since  $\rho \propto \mu$, while
the first term dominates away from half filling (the  Fermi energy $\mu$ is zero at half-filling).
The exponent $\alpha$ contains  $N_c=4$ (8 if  spin included) 
in contrast with  the Fermi liquid case of $N_s  = 2$.  
Since the phase shifts $\delta_\pm$  depends on   $\rho $, the  exponent $\alpha$
also depends  on the Fermi energy  $\mu$, see Fig.\ref{expo}.
In graphene the electron density can be varied up to $10^{13}\mathrm{cm}^{-2}$ by controlling the gate potentials,\cite{exp1,exp2,exp3} and
this opens up the possibility of the systematic measurement of the dependence of the exponent $\alpha$ on the electron density.
As we discuss in  Sec. VII  our result for  the X-ray absorption intensity has several other interesting 
experimental consequences.

This paper is organized as follows:
In Sec. II we describe our model Hamiltonian.
In Sec. III its local action is formulated within  path integral formalism.
The Green functions of Dirac fermion and the core hole are derived in Secs. IV and V.
Using these Green functions we derive the X-ray absorption intensity in Sec. VI.
The discussions are given in Sec. VII.

\section{Model Hamiltonian for graphene}

Graphene can be described by the following tight-binding Hamiltonian of $\pi$-bands.\cite{carbonbook}
\be
\label{tight}
\mathcal{H}_0 = -t \sum_{\mathbf{R},\mathbf{r}_i}\, 
\Big[ c^\dag_A(\mathbf{R}) c_B(\mathbf{R}+\mathbf{r}_i) + \text{h.c.} \Big],
\ee
where $A,B$ denote  sublattices, and ${\bf R}$ (with $\mathbf{r}_i$) spans a honeycomb lattice.
 The hopping parameter is $t= -3.033 \, \mathrm{eV}$, and the basis vectors are given by
$\mathbf{a}_1 = \ell ( \frac{\sqrt{3}}{2}, -\frac{1}{2} ), \;
\mathbf{a}_2 = \ell (0, 1 )$. The lattice spacing is $\ell = 2.46 \text{\AA} $.
Each site of A sublattice is connected to three B sublattice sites:
$\mathbf{r}_1 = \ell ( \frac{1}{2\sqrt{3}}, \frac{1}{2} ),\;\;
\mathbf{r}_2 = \ell ( \frac{1}{2\sqrt{3}}, -\frac{1}{2} ),\;\;
\mathbf{r}_3 = \ell (-\frac{1}{\sqrt{3}}, 0 )$.
The energy eigenvalue is given by
$E_\pm(\vk) = \pm t \vert \Lambda (\vk) \vert$,
where $\Lambda(\vk) \equiv e^{i \vk \cdot {\bf r}_1} + e^{i \vk \cdot {\bf r}_2} +e^{i \vk \cdot {\bf r}_3}$.
Two bands touch at 6 points located along the corners of Brillouin zone, and among them only two are distinct:
$\mathbf{K}_+  = \frac{4\pi }{\sqrt{3} \ell }\, ( \frac{1}{2}, \frac{1}{2 \sqrt{3}} )$ and $
 \mathbf{K}_- = - \mathbf{K}_+$.  These two points will be denoted as  $K$ and $K'$, respectively. 
From now on we will set the lattice constant  $\ell=1$ and $\hbar=1$.

At low energy the above Hamiltonian reduces to the effective Hamiltonian of the massless Dirac fermion,
which is valid within a few eV of the $K$ and $K'$ points\cite{Slo,ando1}.
In  a 
4-component notation involving two from sublattice $A,B$ and two from valley indices  $\mathbf{K}_\pm$,  
the Hamiltonian has the form\cite{ando}
\be
\mathsf{H}_0 = \int d^2 x \, \Psi^\dag(x) \Big( v (-i \partial_x) \alpha^x + 
v (-i \partial_y) \alpha^y \Big) \Psi(x),
\ee
where $v \propto t $ is the velocity of Dirac fermion and 
$\Psi = [c_{A+},c_{B+}, c_{A-},c_{B-}]^T$ is the 4-component spinor (spin indices suppressed). The gamma matrices are
$ \alpha^x = \sigma_x \otimes {\rm I}_2, \quad \alpha^y =  \sigma_y\otimes(-\tau_3) , \quad
\beta = \sigma_z \otimes {\rm I}_2 $, and they satisfy the standard gamma matrix algebra 
 $\{\alpha^i, \alpha^j \}= 2 \delta^{ij}$. $\tau_i, \sigma_i$ are 
two independent copies of Pauli matrices acting on the sublattice and the valley space, respectively.

The Matsubara Green function of Dirac fermion with Fermi energy $\mu$ is given by 
\begin{align}
\label{matsu}
G_{ab} (i\epsilon, \vk) &= -\la \Psi_a   (i\epsilon, \mathbf{k}) \Psi^\dag_b  (i\epsilon, \mathbf{k}) \ra \cr
 &= \left ( \frac{1}{i\epsilon  + \mu - v \sum_{j=x,y} k^j \alpha^j} \right )_{ab}\cr
&=\left ( \frac{i\epsilon+ v \sum_j k^j \alpha^j}{(i\epsilon  + \mu)^2 - v^2 \vk^2 } \right )_{ab},
\end{align}
where  $a,b$ are the collective sublattice/valley indices, $i \epsilon$ is the imaginary frequency, and 
$k^j$ is the j-th component of wavevector ${\bf k}$.
In the last line of the above equation the gamma matrix algebra has been used.

In our model the core potential is taken to be isotropic since  the core hole belongs to
the $s$-level.  
Also, it is assumed that the core potential is screened instantly, and this assumption should be  valid\cite{mahan1}
near the  threshold energy $\omega_T^{*}$.
Thus the  deep core hole can be described by 
\be
\mathsf{H}_{{\rm int}} = d^\dag d \, \Psi^\dag(x=0) \tilde{V}
 \Psi(x=0),
\ee
with the scattering potential $\tilde{V}$ given by Eq.(\ref{potential}).
$d$ is the annihilation operator of the deep core electron and its
 Hamiltonian is given by $E_d \,d^\dag d$, where $E_d < 0$ is the core level energy.
The (bare) Matsubara Green function of the core hole is given by
\be
\label{corebare}
D_0(i\epsilon) = \frac{1}{i\epsilon+\mu -E_d}.
\ee

The interaction with X-ray is described by
\be
\label{x-ray}
\mathsf{H}_X = \sum_{a,\vk} \, M_{a,\vk} \Psi_a^\dag(\vk) d \, e^{-i\omega t} + \text{H.C.}.
\ee
$M_{a,\vk}$ is the X-ray matrix element. Usually the matrix element is very weakly dependent on $\vk$
and its dependence will be ignored. The dependence on the index $a$ will be also ignored.
In this paper the interaction between Dirac fermions is not taken into account.

\section{Formulation in terms of local action}
The \textit{local} Matsubara Green function of Dirac fermions of graphene with the Fermi energy $\mu$ is given by
\begin{align}
\label{localgreen}
& G_{ab}(\tau -\tau') \equiv -\la \Psi_a(\vec{x}=0,\tau) \, \Psi^\dag_b (\vec{x}=0,\tau') \ra \cr
&=\frac{1}{\beta}\,\sum_{\epsilon}\int \frac{d^2 \vk }{(2\pi)^2}\,e^{-i\epsilon(\tau -\tau')} 
G_{ab}(i\epsilon,\vk) \cr
 &=\delta_{ab}\,\frac{1}{\beta}\,\sum_{\epsilon}\int \frac{d^2 \vk }{(2\pi)^2}\,
\, 
\frac{e^{-i \epsilon(\tau -\tau')}(i\epsilon+\mu) }
{(i \epsilon +\mu)^2 - v^2 \vk^2}.
\end{align}
Note that in the last line of the above equation, 
the term depending on momentum linearly [ $\alpha^j k^j$, see Eq.(\ref{matsu}) ] vanished upon wavevector integration.
If $ \vert \tau -\tau' \vert \gg \tau_c$, where $\tau_c$ is a short time cutoff, 
Eq.(\ref{localgreen}) becomes at zero temperature
\begin{align}
\label{localgreen1}
 G_{ab}(\tau -\tau') & =\delta_{ab} G=\delta_{ab} \big( G_{{\rm FL}} +  G_{{\rm D}} \big ),  \cr
 G_{\mathrm{D}}(\tau -\tau') &= - \frac{1}{4\pi v^2}\frac{1}{(\tau -\tau')^2} +\frac{\theta(\tau' -\tau)}{4\pi v^2}
\frac{e^{-\mu (\tau' -\tau)}}{(\tau -\tau')^2}, \cr
G_{{\rm FL}}(\tau -\tau') &= -\frac{\rho}{\tau -\tau'},
\end{align}
where 
\be
\label{dos}
\rho \equiv  \frac{\mu}{4\pi v^2},
\ee
which plays the  role of Dos at the Fermi energy. Note that it is proportional to $\mu$.
Thus the local Green function consists of the Fermi liquid component $G_{{\rm FL}}$ and the pure Dirac component $G_{\mathrm{D}}$.
In the long time/low energy limit the Dirac component is much less singular than the FL component. 
In fact, the direct Feynman diagram calculations at half-filling show the logarithmic singularities 
originating from $G_{{\rm FL}}$  do not exist, so that the perturbation theory
 converges well at low energy. For details, see Sec. VII.

To obtain the action of the local degrees of freedom we have to
 integrate out the bulk degrees of freedom except for the one at $\vec{x}=0$.
For this step,  we use the 
method of Dirac delta function in  \textit{Grassman} variable in the following form
\begin{align}
1&= \int D[\eta_a, \bar{\eta}_a]\, \prod_{a=1}^4 \, 
\delta( \eta_a - \Psi_{a,x=0} )\, \delta(\bar{\eta}_a - \Psi^\dag_{a,x =0}) \cr
&=\int D[\eta_a, \bar{\eta}_a, \lambda_a, \bar{\lambda}_a]\, \exp\Big [ \sum_a \, \bar{\lambda}_a(\tau) ( \eta_a(\tau) - \Psi_a(0,\tau)) \cr
&+ (\bar{\eta}_a (\tau)- \Psi^\dag_a(0,\tau) ) \lambda_a(\tau) \Big ],
\end{align}
where $\lambda_a$ is a Grassman Lagrangian multiplier.
The local action of the local degrees of freedom 
$\eta_a(\tau), \bar{\eta}_a(\tau)$ can be obtained by integrating out
$\Psi_a$ and $\lambda_a$ successively.

The resulting local action of $\eta_a$ and $d$ in imaginary time is 
\begin{align}
\label{action}
&S[\eta,d] = -\sum_{a,b} \int d\tau d\tau' \, \bar{\eta}_a(\tau) 
G^{-1}_{ab}(\tau-\tau') \eta_b(\tau') \cr
&+ \int d \tau \, \bar{d}(\tau) d(\tau) \bar{\eta}(\tau) \tilde{V}
 \eta (\tau)+\int d \tau\,\bar{d}\,(\partial_\tau -\omega_T)  d,
\end{align}
where $\omega_T = \mu -E_d$ is the (bare) threshold energy and
 $G_{ab}^{-1}$ is the inverse of  the local Green function  Eq.(\ref{localgreen}) of the Dirac fermion.
The local action Eq.(\ref{action}) is of a very general form, so that it can be applied to other cases with different
band structures, including FL. The  specific band structure is reflected only in the local Green function
$G_{ab}(\tau-\tau')$.

\section{ The  Green function of Dirac fermion}
In the presence of a following {\it time-dependent} potential
\be
\widehat{V} (\tau) = - \tilde{V} \theta(\tau_1 -\tau) \theta(\tau -\tau_2),
\ee
 the  local Matsubara Green function of Dirac fermions is given by ($\xi,\xi'$ are imaginary times)
\begin{align}
\label{matrixGreen}
 g(\xi -\xi')= \la \xi \vert \frac{1}{G^{-1} - \widehat{V} } \vert \xi' \ra,
\end{align}
which is a 4x4 matrix Green function, and this Green function encapsulates the excitonic contribution to the 
X-ray response function.
From the above definition of $g$ one can show that it satisfies the following (matrix) integral equation.
\be
\label{equation}
g(\xi,\xi') = G(\xi-\xi') + \int_{\tau_2}^{\tau_1} d\tau \,  G(\xi-\tau) (-\tilde{V}) g(\tau, \xi'),
\ee
where $G$ is the bare local Green function, Eq.(\ref{localgreen1}), which is diagonal.
To solve Eq.(\ref{equation}) we choose an ansatz 
\be
\label{ansatz}
g = g_0 {\rm I}_4 + g_1 \sigma_x \otimes \tau_x,
\ee
and take the combinations of $g_0 \pm g_1$. Then it can be shown that each of $g_0 \pm g_1$ satisfies the following equation 
\be
\label{valley3}
(g_{0} \pm g_{1}) = G + (-V_0\mp V_1) \int^{\tau_1}_{\tau_2} d \tau \, G(\xi-\tau) (g_{0} \pm g_{1}).
\ee
When $G$ is of the FL form  $G_{{\rm FL}}(\tau)=-\rho/\tau$,  the solution of Eq.(\ref{valley3})
which  is asymptotically exact \textit{in the long time limit} was 
 obtained by  Nozi\`eres and De Dominicis (ND) \cite{nozieres}:
\begin{align}
\label{ND}
(g_0 \pm g_1)^{{\rm ND}}(\xi, \xi') &= [\cos^2 \delta_\pm] G_{{\rm FL}} (\xi -\xi') \cr 
&\times \left [ \frac{(\xi - \tau_2) ( \tau_1 - \xi') }{
(\tau_1 - \xi)(\xi' - \tau_2)} \right ]^{ \frac{\delta_\pm}{\pi}},
\end{align}
with the 
scattering phase shift  given by
\be
\delta_\pm  = \tan^{-1} [ \pi (V_0 \pm V_1) \rho ].
\ee
This result cannot be obtained perturbatively.
In view of the fact $G = G_{{\rm FL}}+ G_{{\rm D}}$, the Green functions $g_{0,1}$ can be also written 
as a sum of 
\be
 g_{0,1} = g^{{\rm ND}}_{0,1} + \delta g_{0,1},
\ee
where 
$\delta g_{0,1}$ can be computed
perturbatively
since $G_{{\rm D}}$ is much less singular than $G_{{\rm FL}}$ in the long time limit.
Using the above form of $g_{0,1}$  in Eq.(\ref{valley3}) and carrying out the integrals  carefully we find 
that
$\delta g_{0,1}\approx A_{0,1} g^{{\rm ND}}_{0,1}+ B_{0,1} G_{{\rm D}}$ where $A_{0,1}$ are  small multiplicative  
renormalization constants of the order of $ V_{0,1}/t$ and $B_{0,1}$ are also   multiplicative
renormalization constants.
We define, for convenience,  new Green functions $ g^{{\rm FL}}_{0,1}=g^{{\rm ND}}_{0,1}+ A_{0,1} g^{{\rm ND}}_{0,1} $, 
and $g^{{\rm D}}_{0,1}=B_{0,1} G_{{\rm D}}$. 
Note $g^{{\rm FL}}_{0,1}$ vanish as $\mu \to 0$ since they are proportional to $G_{\mathrm{FL}}$, while 
 $g^{{\rm D}}_{0,1}$ do not.
The explicit forms of these  new Green functions are
\begin{align}
\label{result2}
 g^{{\rm FL}}_{0,1}(\tau_1 -\tau_2) & \sim \frac{1}{2} \frac{(-\rho) \cos^2 \delta_+}{\tau_1 -\tau_2}\,
\left ( \frac{ \tau_1 - \tau_2}{\tau_c} \right )^{\frac{2\delta_+}{ \pi}} \cr
&\pm \frac{1}{2}\frac{(-\rho) \cos^2 \delta_-}{\tau_1 -\tau_2}\,
\left ( \frac{ \tau_1 - \tau_2}{\tau_c} \right )^{\frac{2\delta_-}{ \pi}}, \cr
 g^{{\rm D}}_{0}(\tau_1-\tau_2)&\sim -\frac{1}{4\pi v^2} \frac{1}{(\tau_1 -\tau_2)^2}
(1+ \pi
\frac{V_0}{4\pi v^2} \frac{1}{\tau_c}), \cr
g^{{\rm D}}_{1}(\tau_1-\tau_2)&\sim-\frac{1}{4\pi v^2} \left [ \pi
\frac{V_1}{4\pi v^2} \frac{1}{\tau_c} \right ] \frac{1}{(\tau_1 -\tau_2)^2}, 
 \end{align}
where $\rho$ is to be understood as the renormalized Dos at the Fermi energy.
From these it follows that  
\be
  g_{0,1}\approx g^{{\rm FL}}_{0,1}+g^{{\rm D}}_{0,1}.
\ee

%%%%%%%%%%%%%%%%%%%%%%%%%%%%%%%%%%%%%%%%%%%%%%%%%%%%%%%%%%%%%%%%%%%%%%%%%%%%%%%%%%%%%%%%%%%%%

\section{ Core-hole Green functions}
The photoemission spectra is related to the core hole Green function which captures the physics
of OC.
\be
D(\tau_1,\tau_2) =  \la T_\tau \bar{d}(\tau_1) d(\tau_2) \ra.
\ee
At  temperature much lower than
$\omega_T$ this Green function is non-vanishing only for $\tau_1 > \tau_2$. 
Carrying out the path integral over $d$ in 
Eq.(\ref{action}) we obtain 
\be
D(\tau_1,\tau_2) =\frac{ \int D[\eta]\,e^{-S_\eta }
  \la \tau_2 \vert \frac{1}{\partial_\tau  + \bar{\eta} \tilde{V} \eta} \vert \tau_1 \ra}{
\int D[\eta]\,e^{- S_\eta }},
\ee
where 
$ S_\eta $ is the first term of Eq.(\ref{action}).
In fact, a  term $\frac{\tilde{V}}{2}  \delta(\tau -\tau') $ should have been included in $S_\eta$. 
It stems from the determinant of $d$-path integral \cite{kleinert}, but 
it turns out not to affect the long-time behavior, so has been dropped.
$\la \tau_2 \vert \frac{1}{\partial_\tau  +  \bar{\eta} \tilde{V} \eta} \vert \tau_1 \ra$ is 
 Green function (1st order in time)
 in the presence of external potential $ \bar{\eta}  \tilde{V} \eta$, which is easily calculated to be
$  \exp[-\int_{\tau_2}^{\tau_1} d \tau'  \bar{\eta} (-\tilde{V}) \eta ]$.
Now the path integral over $\eta$ can be done.
Recalling $G=G_{{\rm FL}} + G_{{\rm D}}$, the result of the path integral can be expressed as
\be
\label{holegreen}
D(\tau_1,\tau_2)=e^{{\rm Tr} \ln (1 - G_{{\rm FL}}\widehat{V})} e^{{\rm Tr}\ln(1  - \frac{ G_{{\rm D}} 
\widehat{V}}{1 - G_{{\rm FL}}\widehat{V}} )}.
\ee
The first factor of Eq.(\ref{holegreen}) is the core hole Green function of the Fermi liquid component, and this can be calculated 
exactly by using the ND solution Eq.(\ref{ND}).\cite{nozieres,bosobook}
The second factor does not generate  any infrared singularities and 
it can be evaluated by a perturbative expansion in $\widehat{V}$.
Expanding the second factor up to the second order in $\widehat{V}$ we obtain 
\begin{align}
\label{result1}
D(\tau_1, \tau_2)&  \sim e^{-\omega_T^*(\tau_1 -\tau_2)}\,
 \frac{1}{(\frac{\tau_1 -\tau_2}{\tau_c})^{N_c (\delta/\pi)^2}} \cr
&\times 
\exp \Big[- \frac{N_c}{4} \left( \frac{V_0^2+V_1^2}{(4\pi v^2)^2} \right )
 \frac{1}{(\tau_1 -\tau_2)^2}\Big],
\end{align}
where $\omega_T^*$ is the renormalized threshold energy \cite{mahan1} and   $\delta$ is the exponent given in  Eq.(\ref{exp2}).
$N_c=4$ is the number of the species of Dirac fermions, and its presence originates  from the trace over spinor space.
The result Eq.(\ref{result1}) depends critically on band filling. 
At half-filling we have $\delta=0$, and  the correction from the scattering is negligible in the long time limit. 
But away from the half-filling the power law dependence dominates.
Eq.(\ref{result1}) also implies that the overlap between ground state wavefunctions with and without the core hole
is given by
\be
\vert \la 0 \vert V \ra \vert \sim e^{ + \frac{1}{2} N_c (\frac{\delta }{\pi})^2 \ln ( \omega_{{\rm min}} \tau_c)
-{\rm const.} \, ( \omega_{{\rm min}} \tau_c)^2},
\ee
where $\omega_{{\rm min}}$ is the $\pi v /L$ with  the system size $L$. 
At half-filling  the first term in the exponent, which is of the FL type, vanishes since $\delta=0$. However,
 the second term is finite and the overlap becomes
almost independent of the system size. This is is  consistent with the result of Ref.[\onlinecite{guinea07}] 
which was obtained numerically.

The hole spectral function can be  obtained from ($\zeta$ is an infinitesimally small positive number)
\be
\label{defi}
A_h(\omega) = -\mathrm{Im} \Big [ \int d \tau  e^{i\omega \tau} D(\tau)  \vert_{i\omega \to \omega + i \zeta} \Big ].
\ee
In the long time limit (or in the vicinity of  $\omega = \omega_T^*$),  
\begin{align}
\label{hole-spectral2}
& A_h(\omega) \sim \theta(\omega -\omega_T^*)\,\sin [ \pi N_c (\delta/\pi)^2]  
\Big( \frac{ \tau_c c_1}{ [(\omega - \omega_T^*)\tau_c]^{1- N_c (\frac{\delta}{\pi})^2 }} \cr
&- c_2 \tau_c^{-1} [(\omega - \omega_T^*) \tau_c]^{1+ N_c (\frac{\delta}{\pi})^2} \Big ),
\end{align}
where ($\Gamma(x)$ is the Gamma function)
\begin{align}c_1 &=  \Gamma(1-N_c (\delta/\pi)^2 ), \cr
c_2 &= \left(\frac{ V_0^2 + V_1^2}{(4\pi v^2)^2} \right ) 
\frac{\frac{N_c}{4} \,\Gamma(1-N_c (\delta/\pi)^2 )}{(1+N_c (\delta/\pi)^2 ) N_c (\delta/\pi)^2}.
\end{align}
The above expressions of $c_{1,2}$ are valid as long as 
$1 > N_c (\delta/\pi)^2$. 
If $ N_c (\delta/\pi)^2 \ge 1$, the $\tau$-integral of Eq.(\ref{defi}) diverges in the short time 
limit $\tau \to 0$. However, we have to remember that the expression Eq.(\ref{result1}) of $D(\tau)$ is valid 
asymptotically only in the long time limit. Therefore, this divergence is not physically relevant, and it can
be treated by introducing an explicit short time cutoff in the integral or by an
analytic continuation of Gamma functions appearing in $c_{1}$ and $c_{2}$.

At  half-filling we have  $\delta=0$, and  the core-hole Green function  Eq.(\ref{result1}) becomes
\be
\label{duug}
D(\tau_1,\tau_2) \sim e^{-\omega_T^* (\tau_1 -\tau_2)}\, 
\exp \Big[- \frac{N_c}{4} \left( \frac{V_0^2+V_1^2}{(4\pi v^2)^2} \right )
 \frac{1}{(\tau_1 -\tau_2)^2}\Big].
\ee
Expanding the exponent of Eq.(\ref{duug}) in the long time limit we obtain
\begin{align}
A_{h, \delta =0}(\omega) &\sim \pi \delta(\omega-\omega_T^*)    \cr
&-
\theta(\omega -\omega_T^*)  (\omega -\omega_T^*)  \left(\frac{ V_0^2 + V_1^2}{(4\pi v^2)^2} \right ).
\end{align}
Comparing with the result of Eq.(\ref{hole-spectral2}) ($\delta \neq 0$), we find that
the anomalous power-law dependence has disappeared.

%%%%%%%%%%%%%%%%%%%%%%%%%%%%%%%%%%%%%%%%%%%%%%%%%%%%%%%%%%%%%%%%%%%%%%%%%%%
\section{X-ray response function} This is given by
\be
\label{XrayF}
F(\tau_1,\tau_2) =\sum_{a,b} \la T_\tau \bar{d}(\tau_1) \eta_a(\tau_1) 
 \bar{\eta}_b(\tau_2) d(\tau_2)   \ra,  
\ee
which is also non-vanishing only for $\tau_1 > \tau_2$.
To compute this response function it is convenient to introduce the external source $J_a$ for $\eta_{a}$,
$S_J = -\int d\tau \Big(  \sum_a \bar{J}_a \eta_a + \bar{\eta}_a J_a\Big)$, into the local action.
Carrying out $d$-integration and 
$J$-differentiation we arrive at
\begin{align}
F(\tau_1,\tau_2)  &= \sum_{a,b}\frac{\frac{\delta }{\delta J_b (\tau_2)} 
\frac{\delta }{\delta \bar{J}_a (\tau_1)} \int D[\eta]\,e^{-S[\eta] - S_J-
\int d\tau \bar{\eta} \widehat{V} \eta }}{ \int D[\eta] e^{-S[\eta]}} \Bigg \vert_{J \to 0}\cr
&=(-1)\frac{\mathrm{Det}[G^{-1} -\widehat{V}]}{\mathrm{Det}[G^{-1}]}\,\sum_{a,b} g_{ab}(\tau_1 -\tau_2),
\end{align}
where $g_{ab}(\tau_1 -\tau_2)$ is the (matrix) local Green function of the Dirac fermion of Eq.(\ref{matrixGreen}).
We also note that $\frac{\mathrm{Det}[G^{-1} -\widehat{V}]}{\mathrm{Det}[G^{-1}]}$
is the core hole Green function of Eq.(\ref{holegreen}).
The path integral formulation automatically yields that
the excitonic effect (represented by $g_{ab}$) and the OC (represented by the core hole Green function) contribute
in multiplicative way.
Therefore, we obtain
\be
\label{xray2}
F(\tau_1 -\tau_2) = -\theta(\tau_1 -\tau_2)\,D(\tau_1 - \tau_2)\,\sum_{a,b} g_{ab}(\tau_1 -\tau_2).
\ee
From Eq.(\ref{ansatz}) it follows that
\begin{align}
\label{summed}
\sum_{a,b} g_{ab}(\tau_1 -\tau_2) &= 4 (g_0 + g_1) \cr
&=4 \Big[ (g_0^{\mathrm{FL}} + g_1^{\mathrm{FL}} ) + (g_0^{\mathrm{D}} + g_1^{\mathrm{D}} ) \Big ].
\end{align}
Now the absorption  intensity at $T=0$ can be computed from
\be
I(\omega) =-\frac{1}{\pi}\mathrm{Im}\left[ \int_0^\infty\, d \tau \, e^{i \omega \tau}\,F(\tau) 
\Big \vert_{i \omega \to \omega + i \zeta} \right ].
\ee 
Using Eqs.(\ref{result2},\ref{result1},\ref{xray2}) we find for the absorption rate  
\begin{align}
\label{result3}
I= I_{{\rm FL}} + I_{{\rm D}}, 
\end{align}
where
\begin{align}
&I_{{\rm FL}}(\omega) \sim d_{{\rm FL}} \,\theta(\omega -\omega_T^*)\, \rho \, 
\left(\frac{\omega - \omega_T^*}{E_c} \right) ^{-2 \frac{\delta_+}{\pi} + N_c (\delta/\pi)^2}, \nonumber\\ 
&I_{{\rm D}}(\omega)\sim  d_{{\rm D}} \, \theta(\omega -\omega_T^*)\,
\frac{\rho E_c}{\mu}\,\left(\frac{\omega -\omega_T^*}{E_c} \right)^{1+N_c (\delta/\pi)^2},
\end{align}
with the coefficients 
\begin{align}
d_{{\rm FL}} &=  \cos^2 \delta_+ \, \Gamma( 2 \frac{\delta_+}{\pi} - N_c (\delta/\pi)^2 )\,
\sin \big[ \pi ( 2 \frac{\delta_+}{\pi} - N_c (\delta/\pi)^2) \big ], \cr
d_{{\rm D}} &= \sin \big[ N_c ( \delta/\pi )^2 \big ]\,
\frac{\Gamma(1-N_c (\delta/\pi)^2 )}{[1+N_c (\delta/\pi)^2 ] N_c (\delta/\pi)^2}.
\end{align}
The above expressions of $d_{{\rm FL}}$ and $d_{{\rm D}} $ are valid as long as 
$1 > N_c (\delta/\pi)^2$. In the case of $1 \le N_c (\delta/\pi)^2$, $d_{{\rm FL},{\rm D}}$
are given by an analytic continuation of Gamma functions as in the case of the core-hole spectral function.
In particular, $d_{{\rm FL}}$ is positive even if 
$2 \frac{\delta_+}{\pi} - N_c (\frac{\delta}{\pi})^2$ becomes negative,  owing to the identity
 $1= \Gamma(1-x) \Gamma(x) \sin(\pi x)/\pi$.
Note that only the exponent $\delta_+$ appears in Eq.(\ref{result3}) because the second term 
of $g_{0,1}^{\mathrm{FL}}(\tau_1-\tau_2)$ which has a $\delta_-$ exponent [see Eq.(\ref{result2},
\ref{summed})] is canceled in the summation over channels $a,b$ of Eq.(\ref{xray2}).
In the above we have assumed that the X-ray matrix element is very weakly dependent on the sublattice
and valley structure. 
Let us compare our result of the exponent of Eq.(\ref{result3}) with that of Ref.[\onlinecite{larkin}]. 
Eq.(9) of  Ref.[\onlinecite{larkin}] is the \textit{leading term} of the exact result 
Eq.(72) of Ref.[\onlinecite{nozieres}].
In our case  only  $s$-wave {\it orbital} scattering is relevant, and the effect of channels other than
the orbital such as valley and sublattice is reflected in the factor $N_c$. Therefore, our result is 
fully consistent with that of Ref.[\onlinecite{larkin}].

The $I_{{\rm FL}}(\omega)$  vanishes as half-filling is approached 
since $\rho$ and $\delta_\pm \to 0$ at the same time,
 while the $I_{{\rm D}}$ part remains finite.
The exponent  $-2 \frac{\delta_+}{\pi} + N_c (\frac{\delta}{\pi})^2$ 
is a sum of the exponents for the  excitonic and OC processes,  $ -2 \frac{\delta_+}{\pi}$ and 
$N_c (\frac{\delta}{\pi})^2$, respectively.
Away from half-filling, if this exponent is negative, $I_{{\rm FL}}$ is the dominant term since 
it diverges as $\omega \to \omega_T$, manifesting the X-ray singularity.
The $I_{{\rm D}}$  is finite and does not exhibit a singular behavior.

At half-filling,  even in the absence of the local potential, the 
intensity is suppressed by a factor of $\omega -\omega_T^*$ near threshold, and this is a consequence of  the effect of 
the linear Dos. The linear Dos completely suppresses the excitonic processes near the threshold, and the perturbative 
treatment is sufficient, as can be checked explicitly by a direct calculation of Feynman diagrams (see below).

\section{Discussions}

The singular nature of X-ray edge problem of FL can be traced back to the logarithmic divergence of 1-loop
particle-hole polarization function.\cite{mahan1} Due to these logarithmic divergences the infinitely many Feynman
diagrams should be summed even in the weak coupling region, and the summation leads to the anomalous power law
dependence of response functions. Thus to understand the \textit{non-singular} nature of response functions of graphene
\textit{at half filling} ($\mu =0, \rho=0$), it suffices to calculate the 1-loop
particle-hole polarization function and to show that it is non-singular.
The 1-loop polarization function of graphene (at half-filling and at zero temperature) is given by
\begin{align}
&\Pi^{(0)}(i\omega) = -\int \frac{d \epsilon}{2\pi}\,\int \frac{ d^2 \vk}{(2\pi)^2}\,
\mathrm{Tr} \widehat{G}(i\epsilon,\vk) \,  D_0 (i\epsilon -i \omega) \cr
&=-N_c \int \frac{d \epsilon}{2\pi}\,\int \frac{ d^2 \vk}{(2\pi)^2}
\frac{i\epsilon}{(i\epsilon)^2 - v^2 \vk^2} \frac{1}{i\epsilon - i\omega +\omega_T},
\end{align}
where $\widehat{G}(i\epsilon,\vk)$ is the matrix Green function of Eq.(\ref{matsu}) at $\mu=0$, and 
$D_0 (i\epsilon -i \omega)$ is the \textit{bare} core hole Green function, given in Eq.(\ref{corebare}).
In the last line of the above equation the term linear in $\vk$ vanishes upon $\vk$ integration.
The integral is straightforward, and we obtain
\be
\label{bubble}
\Pi^{(0)}(i\omega) \sim \frac{1}{2\pi v^2}\Big[ (i\omega - \omega_T)
\ln \left ( \frac{ \omega_T - i\omega}{E_c}\right ) -E_c  \Big ],
\ee
where $E_c$ is an energy cutoff of the order $t$. As demonstrated in the above (after analytic continuation
$i\omega \to \omega + i \zeta$), the logarithmic singularity near the threshold $\ln (\omega_T- \omega)$ 
is completely suppressed by the prefactor $\omega - \omega_T$ which stems from the linearly vanishing
density of states of graphene. In the case of FL, the prefactor  $\omega - \omega_T$ is replaced by the (finite)
density of states $\rho$. Therefore, for graphene at half-filling,  the higher order corrections are not necessary if we are only interested in the vicinity of threshold.

It would be interesting to observe in X-ray experiments of a single layer graphene\cite{rol,bos}
how the  exponent $\alpha$ of  Eq.(\ref{Noz2}) depends on the electron density.
In graphene the carrier density can be varied up to  $10^{13} \mathrm{cm}^{-2}$ by changing external gates \cite{exp1,exp2,exp3}.
This corresponds to a change in 
the Fermi  energy up to a few 
eV.  The typical  X-ray energy range is given by the energy difference between the core electron in the  $1s$ level 
and Dirac fermions residing  in the $\pi$ band, which 
is in the region of $280 \sim 290 \, \mathrm{eV}$. \cite{carbonbook,rol}
One of the key experimental features would be that 
the absorption spectrum diverges at the threshold energy $\omega_T^{*}$, which would correspond to
a negative exponent $\alpha$ in the first term of Eq.(\ref{Noz2}) (the 
second term of Eq.(\ref{Noz2}) is zero at the at the threshold energy $\omega=\omega_T^{*}$).
In the Fermi liquid the exponent decreases monotonically with the electron density, while in graphene 
it decreases initially and then increases and becomes positive, see Fig.1.  
When the exponent becomes positive the absorption spectrum does not diverge at the threshold.
Experimental verification of this change of sign of the exponent 
in graphene would be intersting.
The experimental measurement of $\alpha$ would also provide useful information about the strength of intra and inter 
valley scattering strengths $V_0$ and $V_1$.

In this paper, the hole was assumed to be immobile (or of infinite mass).\cite{PattonGeller,sham,mass1,mass2}
The finite mass effect and the finite temperature \cite{ohtaka} are expected to smear out the singular 
behavior to some extent, and these factors are to be taken into account for the detailed comparison with experimental data.

\begin{acknowledgements}
This work was supported  by the Grant No. R01-2005-000-10352-0 from the Basic Research Program of the Korean Science and Engineering Foundation,  by the Korea Research Foundation Grant funded by the Korean Government (MOEHRD)(KRF-2005-070-C0044), and by The Second Brain Korea 21 Project.
\end{acknowledgements}
%%%%%%%%%%%%%%%%%%%%%%%%%%%%%%%%%%%%%%%%%%%%%%%%%%%%%%%%%%%

%%%%%%%%%%%%%%%%%%%%%%%%%%%%%%%%%%%%%%%%%%%%%%%%%%%%%%%%%%%%

\begin{thebibliography}{}
\bibitem{comment} In Ref.[\onlinecite{nozieres}]  the short-time correction to $\delta$ [Eq.(44)] were also included 
in the form of  $ \tan \delta = \pi \rho V /(1- \pi \rho V \tan \theta)$. In our paper we will
neglect such short-time correction as $ \pi \rho V \tan \theta$.
\end{thebibliography}
\end{document}